# A New Non Linear, Time Stamped & Feed Back Model Based Encryption Mechanism with Acknowledgement Support

**A.V.N.Krishna**
Indur Institute of Engg. & Tech.,
Siddipet,A.P., India.
**hari_avn@rediffmail.com**

**A.Vinaya Babu**
J.N.T.U.H, Hyderabad.

*Abstract*

In this work a model is going to be used which develops data distributed over a identified value which is used as nonce (IV). The model considers an equilibrium equation which is a function of non linear relationships, time variant and nonce variant values and takes the feed back of earlier round as input to the present round. The process is repeated for different timings which are used as time stamps in the encryption mechanism. Thus this model generates a distributed sequence which is used as sub key. This model supports very important parameters in symmetric data encryption schemes like non linear relationships between different values used in the model, variable key length, timeliness of encryption mechanism and also acknowledgement between the participating parties. It also supports feed back mode which provides necessary strength against crypto analysis.

## 1. Introduction

Historically, encryption schemes were the first central area of interest in cryptography[1-9]. They deal with providing means to enable private communication over an insecure channel. A sender wishes to transmit information to a receiver over an insecure channel that is a channel which may be tapped by an adversary. Thus, the information to be communicated, which we call the plaintext, must be transformed (encrypted)to a cipher text, a form not legible by anybody other than the intended receiver. The latter must be given some way to decrypt the cipher text, i.e. retrieve the original message, while this must not be possible for an adversary. This is where keys come into play; the receiver is considered to have a key at his disposal, enabling him to recover the actual message, a fact that distinguishes him from any adversary. An encryption scheme consists of three algorithms: The encryption algorithm transforms plaintexts into cipher texts while the decryption algorithm converts cipher texts back into plaintexts. A third algorithm, called the key generator, creates pairs of keys: an encryption key, input to the encryption algorithm, and a related decryption key needed to decrypt. This work mainly deals with the algorithm which generates sub keys which provides sufficient strength to the encryption mechanism.

Partial differential equations to model multiscale phenomena are ubiquitous in industrial applications and their numerical solution is an outstanding challenge within the field of scientific computing[10-12]. The approach is to process the mathematical model at the level of the equations, before discretization, either removing non-essential small scales when possible, or exploiting special features of the small scales such as self-similarity or scale separation to formulate more tractable computational problems. Types of data are as follows:

- *Static:* Each data item is considered free from any temporal reference and the inferences that can be derived from this data are also free of any temporal aspects
- *Sequence:* In this category of data, though there may not be any explicit reference to time, there exists a sort of qualitative temporal relationship between data items.
- *Time stamped:* Here we can not only say that a transaction occurred before another but also the exact temporal distance between the data elements. Also with the events being uniformly spaced on the time scale.
- *Fully Temporal:* In this category, the validity of the data elements is time dependent. The inferences are necessarily temporal in such cases.

## 2. Literature survey

Any symmetric encryption scheme uses a private key for secure data transfer. In their work [19], the authors considered not only key but also time stamp and nonce values to increase the strength of sub key generated. In addition the nonce value can also be used for acknowledgement support between participating parties. The model can be further improved by considering a non linear model where the key values vary with the data generated. In another work [18], the authors considered sub key generations from the master key with the time stamped values which are dynamic in nature. In work [15], the authors considered a random matrix key for sub key generation which is a random number.

## 3. Mathematical modeling of the problem

The approach to time series analysis was the establishment of a mathematical model describing the observed system. Depending on the appropriation of the problem a linear or nonlinear model will be developed. This model can be useful to generate data at different times to map it with plain text to generate cipher text.

### *3.1. Non Linear data flow Problem*
Consider the equilibrium equation

$$T / T_\delta = a + b (u/u_\delta) + c (u/u_\delta)^2$$

Where T is the data to be calculated, $T_\delta$ is a seed value, u & $u_\delta$ are non linear values which vary along with Data T.

If a, b, c are considered as key which are constants, the equilibrium equation will be linear in nature and the values of T calculated are easily susceptible to crypto analysis. To make it strong against crypto analysis, the model is made non linear with varying values for a, b & c, where

$$a = 1 + r \{(\gamma[i] - 1) * M[j]^2\} * (1-\theta)$$

$$b = r \{(\gamma[i] - 1) * M[j]^2\} * (\theta)$$

$$c = - r \{(\gamma[i] - 1) * M[j]^2$$

$$\theta = (T - T_w) / (T - T_\delta)$$

delt = 1.

M[j] = Different time stamps considered for j intervals ,

$\gamma[i]$ = Different space steps considered for i variants

r be the key considered.

To make the relationship between T & u, non linear, a random number is generated which is used for relationship between the two quantities (see Table 1).

**Table 1 : Relationship Values Between T and U**

| T | $T_\delta$ | $T_w$ | U | $u_\delta$ |
|---|---|---|---|---|
| 0-5 | 0.3 T | 0.6 T | 3 | 0.3 u |
| 6-10 | 0.4 T | 0.6 T | 2 | 0.4 u |
| 11-15 | 0.5 T | 0.7 T | 5 | 0.5 u |
| 16-20 | 0.6 T | 0.7 T | 1 | 0.6 u |
| 21-25 | 0.7 T | 0.8 T | 7 | 0.7 u |
| 26-30 | 0.8 T | 0.8 T | 4 | 0.8 u |
| 31-35 | 0.9 T | 0.9 T | 6 | 0.9 u |

*Algorithm*

```
  int r, g,m,t,t1,u,u1;
 float t2,t3,a,b,c,x,y,z;
 t1=4;
 t=t1/0.3;
 read  &r,&g,&m,&u,&u1;
 for(m=1; m<6;m++)
 {
 for(g=2;g<24;g++)
 {
 t2=.6*t;
 t3=(t-t3)/(t-t2);

 x=g-1;
 x=x/2;
 y=pow(m,2);
 z=t3-1;
 a= 1+(r*x*y*z);
 b=t3*r*(x*y);
 c=-r*x*y;
 t=a+b*(u/u1)+c*(pow((u/u1),2));
 printf("%d",t%35 );
 }
 printf("%d", t%35);
 }
```

## 4. Results

By considering a suitable key r =4, for a total time stamp of 5 units, M[6] = { 1,2,3,4,5,6} and γ[i]={ 2,3,4,5,6,7,8,9,10,11,12,13,14,15,16,17,18,19,20,21}. Different data values obtained are

7 5 9 17 25 31 28 15 21 27 33 34 10 16 22 28 34 5 11 17 23 24 5 31 21 10 34 24 13 2 26 15 4 28 17 6 30 19 8 32 21 1 01 4 34 19 30 16 12 04 24 08 27 01 04 33 17 12………….

Thus by using the same key, by changing the time stamp & nonce values different sequences can be generated which are used as sub keys which are of variable in length. These sub keys can be mapped to plain text to generate cipher text [13,15] (see Table 2).

**Table 2: Encryption**

| Plain Text | A | S | k | S |
|---|---|---|---|---|
| Conversion to alpha numeric value | 10 | 28 | 20 | 28 |
| Sub key | 07 | 05 | 9 | 17 |
| Total | 17 | 33 | 29 | 45 |
| Mod 36 | 17 | 33 | 29 | 09 |
| Cipher Text | G | X | t | 09 |

**Table 3: Decryption**

| Cipher Text | G | X | t | 09 |
|---|---|---|---|---|
| Conversion to alpha numeric value | 17 | 33 | 29 | 09 |
| Add 36 if less than 9 | 17 | 33 | 29 | 45 |
| Sub key | 07 | 05 | 09 | 17 |
| Subtract | 10 | 28 | 20 | 28 |
| Plain Text | A | S | k | s |

## 5. Security Analysis

*Analysis by Construction:* In the given model, even though a single valued key is used, it is non linear in nature. By changing the data values, the values of u are also changes as per predetermined values. The relationship between T and is maintained such that the relationship is non linear. The values a, b & c are so designed such that it is not only the key, but also different time stamps, variable length nonce values influence in their generation of values. And also the model uses a feed back mode where generated values are input to next level of data values to be generated. Thus this model provides good security against crypto analysis. The complexity of the algorithm O(n) * O( st) where n is the length of the key, t is the time stamp steps & s be the variable nonce value.

*Complexity by its strength:* Since the model involves not only the key, time stamps, variable nonce values, but also data of past time stamps, it is relatively free from cipher text attack, known plain text & cipher text attacks. A mod function is used on output in each round, it is relatively free from differential crypto analysis. Since a variable length key can be generated, it will also be free from linear crypto analysis.

## 6. Conclusion & Future Work

This encryption mechanism uses an Initialization Vector, Time Stamp & Key to generate distributed sequences which are used as sub-keys. Since the time stamp is variable in nature, the model provides sufficient security against crypto analysis. The model is free from cipher text attack, known plain text & cipher text attacks. In the given model past & present time stamps have been used to generate data. By properly guessing future time stamps, the model can be made still stronger.

## References

[1] Henry Baker and Fred Piper : Cipher systems(North wood books, London 1982).
[2] I.Chien-Chiang: Efficient improvement to XTR and two padding schemes for probabilistic trapdoor one way function, 2005-12-05.
[3] J.William stalling **:**Cryptography and network security  (Pearson Education,ASIA1998)
[4] Krishna A.V.N.: A new algorithm in network security, International Conference Proc. Of CISTM-05, 24-26 July 2005, Gurgoan, India.
[5] Krishna A.V.N., Vishnu Vardhan.B:Decision Support Systems in Improving the performance of rocket Missile systems, Giorgio Ranchi, Anno LXIII,n-5 Septembre-October 2008, pp607-615.
[6] Krishna A.V.N., S.N.N.Pandit: A new Algorithm in Network Security for data transmission, Acharya Nagarjuna International Journal of Mathematics and Information Technology, Vol: 1, No. 2, 2004 pp97-108
[7] Krishna A.V.N, S.N.N.Pandit, A.Vinaya Babu: A generalized scheme for data encryption technique using a randomized matrix key, Journal of Discrete Mathematical Sciences & Cryptography, Vol 10, No. 1, Feb 2007, pp73-81
[8] Krishna A.V.N., A.Vinaya Babu: Web and Network Communication security Algorithms, Journal on Software Engineering, Vol 1,No.1, July 06, pp12-14
[9] Krishna A.V.N, A.Vinaya Babu: Pipeline Data Compression & Encryption Techniques in e-learning environment, Journal of Theoretical and Applied Information Technology, Vol 3, No.1, Jan 2007, pp37-43
[10] Lester S. Hill, Cryptography in an Algebraic Alphabet, The American Mathematical Monthly 36, June-July 1929, pp306–312.
[11] Lester S. Hill, Concerning Certain Linear Transformation Apparatus of Cryptography, The American Mathematical Monthly 38, 1931, pp135–154.
[12] Pandit S.N.N (1963): Some quantitative combinatorial search problems. (Ph.D. Thesis).
[13] Phillip Rogaway : Nonce Based Symmetric Encryption, www.cs.ucdavis.edu/rogeway.
[14] Raja Ramanna Numerical methods 78-85(1990).
[15] Prasad Reddy et al: Data encryption technique using location based key dependent permutation and circular rotations, IJCNS, Vol 2, No.3, March 2010.
[16] R.S.Thore & D.B.Talange: Security of internet to pager E-mail messages (Internet for India-1997IEEE Hyderabad  section) pp.89-94.
[17] Suhas V. Patenkar Numerical Heat Transfer and Fluid Flow 11-75(1991).
[18] Huy  Houng Hyo et al: Dynamic key cryptography & applications, IJNS, Vol 10, No.3, pp 161-174, May 2010.
[19] Krishna A.V.N et al: A New model based encryption scheme with time stamp & acknowledgement support, IJNS, Jan 2011.